\documentclass{kapproc} 
\let\footnote\savefootnote
\let\footnotetext\savefootnotetext 
\let\footnoterule\savefootnoterule 
\kluwerbib

\RequirePackage{graphicx}%
\RequirePackage{epsf}%

\begin{document}

\articletitle{Strong Absorption-line Systems at Low Redshift:  MgII
and Damped Ly$\alpha$}

\author{Daniel B. Nestor, Sandhya M. Rao, David A. Turnshek}

\affil{University of Pittsburgh}
\email{dbn@phyast.pitt.edu}

\chaptitlerunninghead{DLAs @ Low-z.}

\begin{abstract}
We detail a powerful indirect method for the study of damped Ly$\alpha$
systems (DLAs) at low redshift.  We increase the probability of
finding a low-redshift DLA to nearly 50\% by targeting QSOs that are
known to have strong low-redshift MgII and FeII absorption lines in
their spectra.  We are using Sloan Digital Sky Survey QSO spectra
complemented by a survey we are conducting at the MMT
to study the metal-line systems. The Hubble Space Telescope
is being used to confirm low-redshift DLAs.  In addition, we are imaging 
low-redshift DLA galaxies with several ground-based telescopes to 
directly study their environments.
\end{abstract}

\section{Introduction}

Damped Ly$\alpha$ systems (DLAs), QSO absorption line systems with
column densities $N(HI) \ge 2\times 10^{20}$ atoms cm$^{-2}$, contain
less than 10\% of the baryons in the universe. However, since over 95\%
of the neutral gas content of the universe resides in
DLAs, they are extremely important for the study of  galaxy formation
and evolution. In particular, low-redshift DLAs  provide a unique
opportunity to study the gaseous and luminous components of galaxies
simultaneously.

In order to fully understand the evolution of the neutral gas component,
one needs to know {\it how much} of it exists and
{\it where} it exists at every epoch. In the absence of selection effects 
which recent observations suggest are not important (Ellison et al. 2001), 
the answer to the former 
can be found in the statistics of DLAs. The whereabouts of this
neutral gas relative to starlight 
can be studied if the DLA galaxies can be identified.  
Eventually we hope to understand the evolution of DLAs consistently
with other studies of galaxy evolution such as the star formation 
history of the universe, the 
evolution of galaxy number counts, and simulations of structure 
formation and evolution.

There are, however, difficulties inherent to the study of DLAs.  
Specifically, imaging DLAs is difficult at high redshift ($z> 1.65$) 
where the Ly$\alpha$ transition is observed in the optical and the
statistics are better understood, while the statistics are poor at
low redshift where imaging studies are more practical.  Only $\approx
30$ low-redshift DLA absorbers are known from UV
spectroscopic surveys in comparison to the over 100 high-redshift ones
found in optical spectroscopic surveys.  Thus, the low-redshift 
DLA number density, and hence  $\Omega_{DLA}$, 
the cosmological neutral gas mass density in
DLAs, are poorly known (see Figure 1).  
\begin{figure}
\epsscale{0.55}
\plotone{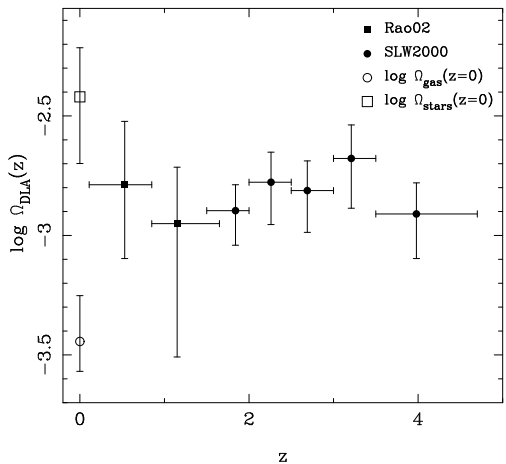}
\caption{The log of $\Omega_{DLA}$ vs. redshift. The two 
low-redshift ($z< 1.65$)
data points are from Rao et al. (2002, in preparation). These include
addition data obtained after the publication of RT2000.}
\end{figure}
Given that the look-back
time associated with redshift 1.65 is $\approx 70\%$ of the age of
the universe, this is a serious limitation.  

DLA galaxies are sometimes elusive even at low
redshift (Steidel et al. 1997, Bouch\'e et al. 2001).  Since only
a  handful of low-redshift DLA galaxies have been identified 
(Le Brun et al. 1997
and Nestor et al. 2001 describe most of them), a detailed study
of DLA environments is not yet possible. 
In this contribution we describe our efforts to
increase the sample of low-redshift DLAs, addressing both the 
{\it how much} and the {\it where} of neutral gas in the universe,
as well as some recent results.

\section{DLAs and Low-Ionization Metal Line Systems}

Rao \& Turnshek (2000, hereafter RT2000) uncovered an empirical
relation  between the strengths of the MgII$\lambda2796$ and
FeII$\lambda2600$  absorption lines and the occurrence of DLAs (Figure
2). They found that  nearly 50\% of systems with
$W_0^{\lambda2796}>0.5$ \AA\ and  $W_0^{\lambda2600}>0.5$ \AA\
were DLAs, and the rest were subDLAs with $10^{19} \le N(HI) < 2\times
10^{20}$ atoms cm$^{-2}$. This discovery has two implications for
low-z   DLA work.  First, targeted surveys for DLAs in QSO spectra
with known  strong low-ionization metal line absorption systems will
have a much  greater success rate.  Second, these metal lines appear
in the optical  down to much lower redshift (for a spectrograph with
coverage down to  3200\AA, for example, FeII$\lambda2600$ can be found
down to $z=0.23$ and  MgII$\lambda2796$ to $z=0.14$.)  Because of the
relative ease of acquiring  optical QSO spectra, $dn/dz$ for these
systems is much better known,  thus allowing us to ``bootstrap'' from
$dn/dz$ for the metal-line systems  to that of DLAs.  We are
undertaking targeted surveys with HST to discover  $z< 1.65$ DLAs in
known strong low-ionization metal line absorption systems.  The goals
are to provide new systems at low enough redshift for imaging, to
better determine the statistical properties of DLAs including $dn/dz$,
$\Omega_{DLA}$, and their column-density distribution, $f(N)$, at
$z< 1.65$ as well as to to better understand the relationship
between FeII absorption, MgII absorption, and the properties of the
Ly$\alpha$  line associated with them.

\begin{figure}
\epsscale{0.55} \plotone{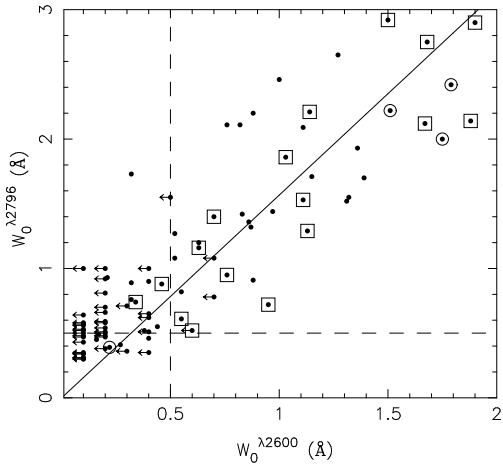}
\caption{$W_0^{\lambda2796}$ vs. $W_0^{\lambda2600}$ for low-redshift
systems with Ly$\alpha$ information. DLAs are represented by squares
around the data point. Circles are DLAs discovered by 21-cm studies.
Nearly 50\% of systems in the upper right quadrant are DLAs.}
\end{figure}

An observationally ideal sample for discovering DLAs would include
bright QSOs with strong intervening metal-line  absorption systems at
redshifts close to that of the QSO.  A low value of
$z_{em}-z_{abs}$ reduces a chance occurrence of an intervening Lyman
limit system that might result in the absence of continuum flux at the
expected wavelength of the DLA line. Thus, the first step is to find
large samples of low-redshift metal-line systems from which
homogeneous subsets  with specific selection criteria can be
constructed. The Sloan Digital Sky Survey (SDSS)  is an excellent
resource for this as it  provides over 3800 optical QSO spectra in the
Early Data Release (EDR, June 2001) alone.

\section{Metal-Line Statistics from the SDSS} 

We have constructed an unbiased sample of 640 absorption line systems
with $W_0^{\lambda2796} > 1.0$ \AA\ from the SDSS EDR.   The task of
analyzing several thousand spectra requires automation and  good
spectral analysis algorithms.  We used our own continuum-fitting and
line-finding routines and interactively confirmed or rejected each
candidate absorption system. We also checked for blending and the
quality of the continuum fit.  Initial tests suggest that the 1.0 \AA\
sample is complete, and analysis of the sample with $W_0^{\lambda2796}
> 0.5$ \AA\ is in progress. Statistical results of the 1.0 \AA\ sample
are discussed in Nestor et al. (these proceedings). We find that there
is evidence for mild evolution with redshift for both $dn/dz$,
especially at $z < 0.6$, and in the slope of the $\log
W_0^{\lambda2796}$ distribution.

The metal-line $dn/dz$ results from the SDSS will greatly reduce the
statistical errors in the calculation of the DLA $dn/dz$  and
$\Omega_{DLA}$ for $z>0.4$. Currently, $\approx$50\% of the error  in
$dn/dz$ for the DLAs comes from the error in $dn/dz$ for the MgII
systems. We are also conducting a large survey for MgII and FeII
absorption  at the  MMT  which will eventually
permit improved estimates of $\Omega_{DLA}$ down to $z=0.14$.

\section{DLA Galaxies} 

\begin{figure}
\epsscale{0.6} \plotone{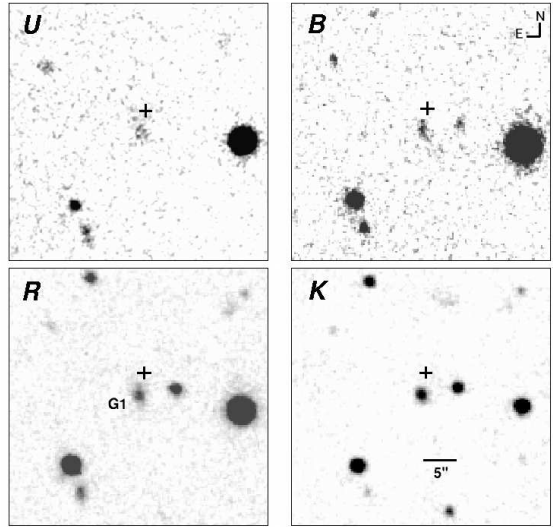}
\caption{UBRK images of the PKS 1629+120 field ($z_{DLA}=0.532$.)
North is up and East is left. The QSO PSF has been subtracted and its
position marked with a ``+''. G1 is identified as the DLA galaxy.}
\end{figure}
\begin{figure}
\epsscale{0.6} \plotone{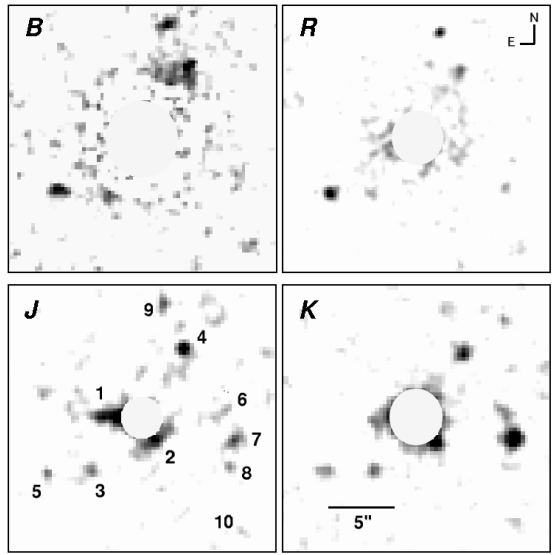}
\caption{Smoothed BRJK images of the PKS 0952+179 field
($z_{DLA}=0.239$.)  North is up and East is left. The QSO PSF has been
subtracted and the residuals masked out. See text for details.}
\end{figure}

The $\approx14$ low-redshift DLA galaxies that have been imaged  thus
far reveal a mix of morphologies ranging from compact dwarf galaxies
to large spirals (Le Brun et al. 1997, Steidel et al. 1995, Turnshek
et al. 2001, Nestor  et al. 2001). There also appear to be a higher
fraction of low surface brightness galaxies among the DLA galaxies in
comparison to the general field.  Here, we present imaging results on
two of the DLA fields. For details  see Rao et al. (2002).

Figure 3 shows UBRK images of the PKS 1629+120  field that has a DLA
at $z=0.532$. The only resolved object within 10\arcsec\ of the QSO 
is a  spiral galaxy (labeled G1) with an impact parameter of $\approx 17$
 kpc.  Although the galaxy redshift has not
been spectroscopically confirmed,  it would be $\approx L^*$ at
$z=z_{abs}$. Spectral evolution synthesis model fits to the photometry
suggest that if the galaxy was at $z=0.532$ it would consist of a
young stellar population with some dust reddening in addition to an
underlying older population.  The UBRK morphologies 
in Figure 3 are consistent with this interpretation. This sightline 
also has a subDLA system at $z=0.901$. G1 would be an unusually luminous,
$\approx 4L^*$, galaxy at this redshift and is therefore 
unlikely to be the $z=0.901$ galaxy. 

The PKS 0952+179 field ($z_{DLA}=0.239$) shown in Figure 4 is quite
different.  Faint disk-like structures, labeled 1 and 2, can be seen
to the immediate east and southwest of the QSO sightline in the J-band
image, which has the best seeing and QSO PSF subtraction of the four.
They span $\approx 25$ kpc. PSF subtraction of the light from the QSO
leaves residuals that are $\approx 0.02L^*$ at $z=0.239$. 
Several additional features within
this $20\arcsec\times 20\arcsec$ field are visible in at least 2 of
the images. The relatively bright object in the K-band image to the
west of the QSO sightline, \#7, has colors that classify it as an
extremely red object (ERO).  If it is   related to the $z=0.239$ DLA
galaxy it would be the lowest redshift  ERO known. The conclusion 
would be that it is a starbursting region with very strong dust
extinction ($A_V\gtrsim 4.8$).

\section{Summary}
Through both indirect and direct methods, we are making  progress
in  the attempt to understand the low-redshift DLA population.
The SDSS and our own MMT survey spectra are enabling us to determine
the statistical properties of low-redshift MgII and FeII absorption systems, 
which can be used to track high $N(HI)$ systems at $z< 1.65$
to a high degree of accuracy.  With our ongoing HST surveys we 
are finding new low-redshift DLAs, improving the low-redshift DLA 
statistics, and improving our  understanding of the empirical relation 
between low-ionization  metal-line systems and DLAs.
In addition, we are directly studying the DLA environment with an active 
imaging campaign.




\begin{chapthebibliography}{99}
\bibitem{B} 
Bouch\'e, N. et al. 2001, ApJ, 550, 585
\bibitem{E}
Ellison, S. L. et al. 2001, A\&A, 379, 393
\bibitem{L}
Le Brun, V. et al. 1997, A\&A, 321, 733
\bibitem{nestor}
Nestor, D. B. et al. 2001, in {\it Extragalactic Gas at Low Redshift}, ASP 
Conf Ser, eds. J. Mulchaey and J. Stocke, p. 34
\bibitem{RT2000}
Rao, S. M., \& Turnshek, D. A. 2000, ApJS, 130, 1 (RT2000)
\bibitem{Retal02}
Rao, S. M., Nestor, D. B., Turnshek, D. A., Monier, E., Lane, W.,
\& Bergeron, J. 2002, in preparation
\bibitem{95}
Steidel, C. C., Bowen, D., Blades, C., \& Dickinson, M., 
1995, ApJ, 440, L45
\bibitem{97}
Steidel, C. C. et al. 1997, ApJ, 480, 568
\bibitem{T}
Turnshek, D. A., et al. 2001, ApJ, 553, 288

\end{chapthebibliography}
\end{document}